\documentclass[10pt]{article}
\usepackage{geometry}
\usepackage{authblk}
\usepackage{graphicx}
\usepackage{amsmath}
\usepackage{multirow}
\usepackage{caption}
\usepackage{varwidth}
\usepackage{enumitem}
\setlist{nolistsep}

\begin{document}

\title{\textbf{Hybrid CMOS-CNFET based NP dynamic Carry Look Ahead Adder}}
\author{A. Nagalakshmi, Ch. Sirisha, Dr. D.N. Madhusudana Rao}
\affil{Department of Electronics and Communication Engineering, Gayatri Vidya Parishad College of Engineering for Women, Visakhapatnam, India.}

\date{Jan 01, 2016}
\maketitle

\begin{abstract}
Advanced electronic device technologies require a faster operation and smaller average power consumption, which are the most important parameters in very large scale integrated circuit design. The conventional Complementary Metal-Oxide Semiconductor (CMOS) technology is limited by the threshold voltage and subthreshold leakage problems in scaling of devices. This leads to failure in adapting it to sub-micron and nanotechnologies. The carbon nanotube (CNT) technology overcomes the threshold voltage and subthreshold leakage problems despite reduction in size. The CNT based technology develops the most promising devices among emerging technologies because it has most of the desired features. Carbon Nanotube Field Effect Transistors (CNFETs) are the novel devices that are expected to sustain the transistor scalability while increasing its performance. Recently, there have been tremendous advances in CNT technology for nanoelectronics applications. CNFETs avoid most of the fundamental limitations and offer several advantages compared to silicon-based technology. Though CNT evolves as a better option to overcome some of the bulk CMOS problems, the CNT itself still immersed with setbacks. The fabrication of carbon nanotube at very large digital circuits on a single substrate is difficult to achieve. Therefore, a hybrid NP dynamic Carry Look Ahead Adder (CLA)  is designed using p-CNFET and n-MOS transistors. Here, the performance of CLA  is evaluated in 8-bit, 16-bit, 32-bit and 64-bit stages with the following four different implementations: silicon MOSFET (Si-MOSFET) domino logic, Si-MOSFET NP dynamic CMOS, carbon nanotube MOSFET (CN-MOSFET) domino logic, and CN-MOSFET NP dynamic CMOS. Finally, a Hybrid CMOS-CNFET based 64-bit NP dynamic CLA is evaluated based on HSPICE simulation in 32nm technology, which effectively suppresses power dissipation without an increase in propagation delay.
\end{abstract}

\section{Introduction}
Nowadays, Very Large Scale Integrated (VLSI) industry requires ultra high speed, low chip area, low power consuming, and portable processors. To fulfil the above requirements, there is a need to scale the devices and device interconnects. This leads to Moore’s law of scaling in integrated circuts. But, CMOS device scaling leads to leakage currents, short channel effects like velocity saturation and mobility degradation which are undesirable. These limits can be overcome to some extent by modifying the channel material in the traditional bulk MOSFET structure with a single CNT or an array of CNTs. For the past three decades the scaling of MOSFET has been the driving force towards the technological advancement. However, for VLSI systems, which depend on Silicon MOS technology, Industry Technology Road Map for Semiconductors (ITRS) has predicted that in nano regimes the expected high density integration will encounter substantial difficulties due to fundamental limitations in physics, material, and manufacturing obstacles [1].

There is a pressing need to explore circuit design ideas in new emerging technologies in deep-submicron regime, in order to exploit their full potential during the early stages of their development. According to the Stanford nano electronics lab[2], CNTs could launch a new generation of faster electronic devices that use less energy than those built using silicon-based transistors. Its electrical properties of greater mobility and high current carrying capability offer the potential for evolving towards the next generation of devices and circuits. Therefore, CNT would be a promising candidate for future nano-scale transistor devices.

\section{Proposed circuit and logic technique}
Dynamic logic circuits are used for high performance and high speed applications. Two different techniques exist in dynamic circuit implementation based on MOS technology. They are the Domino logic circuit technique and NP dynamic CMOS circuit technique. These techniques are employed to satisfy the monotonicity requirement in dynamic circuits [3]. CLA's are commonly used in modern processors. CLA improves speed by reducing the amount of time required to calculate carry bits. The CLA calculates one or more carry bits before the sum, which reduces the wait time to calculate the result of larger value bits. CLA implementation with conventional MOSFET technology in both the techniques prefers domino logic circuit to NP dynamic logic. Comparatively, power wastage is less in conventional MOSFET domino logic circuit than NP dynamic CMOS circuit.

Our objective is to implement CLA up to 64-bit value by replacing the MOS technology with CNT technology in both domino logic and NP dynamic logic circuit techniques. Finally, a hybrid CMOS-CNT based NP dynamic 64-bit CLA is implemented and its performance in terms of MOSFET, CNFET and Hybrid CMOS-CNT technologies is analyzed. 

\section{Implementation of the proposed method}

The proposed methodology is the hybrid CMOS-CNT based NP dynamic CLA implementation. The work is carried out in stages initiating with silicon based CLA. The input of CLA stages are considered as 8-bit, 16-bit, 32-bit and 64-bit. Similar steps are followed by replacing silicon with carbon nanotubes. A 64-bit CLA based on CNT technology is implemented with Stanford specifications. Both the domino and NP dynamic circuits are taken under study and results are observed. In CNT method, NP dynamic CLA functioning improves much than that of domino in terms of power dissipation and delay. Therefore, the NP dynamic logic CLA is considered and hybrid CMOS-CNT working on stages respectively is processed.\\

The CNT specifications are listed below:
\begin{itemize}
\item \textbf{Chirality}: CNFET utilizes a single carbon nano-tube or an array of carbon nano-tubes as the channel material instead of bulk silicon in the traditional MOSFET structure. It was first demonstrated in 1998. The structure of CNT described by an index with a pair of integers (n, m) that define its chiral vector.
\begin{equation}
\begin{split}
\theta=\tan^{-1}\left(\frac{\sqrt{3}n}{2m+n}\right),\\
d_t= \frac{L}{\pi} = \frac{a}{\pi} \sqrt{n^2+nm+m^2},
\end{split}
\end{equation}
where $\theta$ is chiral angle.

For $n = m$; bonding is metallic. For $n–-m = 3I$; it has a small band gap. Whereas, for $n-m \neq 3I$; it is semiconducting. The chiral angle is used to separate carbon nanotubes into three classes differentiated by their electronic properties as follows:
\begin{itemize}
\item armchair for $n = m,~\& ~\theta= 30^o$,
\item zig-zag for $m = 0, n > 0, ~\&~ \theta = 0^o$,
\item chiral for $0 < |m| < n, ~\&~ 0 < \theta < 30^o$.
\end{itemize}

\item \textbf{Channel length ($L_{ch}$)}: The channel length is chosen to reduce the occurrence of scattering. In this paper, we have taken $L_{ch}=32$nm.

\item \textbf{Diameter (D):} The diameter of CNT is between 1.2nm and 1.8nm. In this range, the chirality vectors for zigzag tubes is (16, 0) (17, 0) (19, 0)(20, 0) (22, 0). In this paper, (17, 0) is chosen as the chirality vector.
Diameter of the CNT is given by
\begin{equation}
 D = n \sqrt{3} a_{cc}~,
\end{equation}

where $n$ denotes chirality vector, and $a_{cc}$ is lattice constant which is 0.142nm for carbon. Diameter obtained using Eq.(2) was found to be 1.33nm. 

\item \textbf{Pitch:} It is defined as the minimum distance between two adjacent carbon nanotubes. It can be calculated by the formula

\begin{equation}
\text{Pitch}=\frac{W_g-d}{N-1}~,
\end{equation}

where $W_g$ is the gate width, $d$ is the diameter of the CNT, and N is the no. of parallel channels.

 For our simulation, we have chosen $W_g= 32$nm, $d=1.3$3nm, and $N=9$. 
 
 \item \textbf{Oxide thickness ($t_{ox}$)}: For channel length of $32nm$, $t_{ox}$ is prescribed to be $4nm$.
 
\item \textbf{Dielectric constant ($k_{ox}$):} The dielectric constant for carbon is between 12 and 16. As k increases, power consumption reduces. So, we have taken $k_{ox}=16$.

\item \textbf{Length of doped CNT source-side ($L_{SS}$) and drain side ($L_{DS}$) extension:} The length of source side as well as drain side extension are taken to  same and equal to the channel length, which is 32nm.

\end{itemize}

 

In the Stanford CNFET model, physical channel length is set to be 32nm. The results reported in this paper were obtained using ``\textbf{HSPICE}" 32nm technology[2,4-6].  In comparison to silicon based devices, CNFET provides better control over channel formation, better threshold voltage, better sub-threshold slope, high mobility, high current density, and transconductance.

Implementation of CLA circuit in NP dynamic and Domino logic techniques with increasing 8-bit, 16-bit, 32-bit and finally reaching to 64-bit operation is referred from our previous work [7], which is further more enhanced and corrections are made wherever necessary to carry on with the proposed design. The proposed design integrates carbon nanotube (CNT) fabrication with standard commercial CMOS very large scale integration on a single substrate suitable for emerging hybrid nanotechnology applications. This co-integration combines the inherent advantages of CMOS and CNTs. This hybrid design demonstrates the successful co-integration on a single chip utilizing both silicon n-channel MOSFET and p-type CNT transistors.

\begin{figure}[h!]
\begin{center}
\includegraphics[scale=0.7]{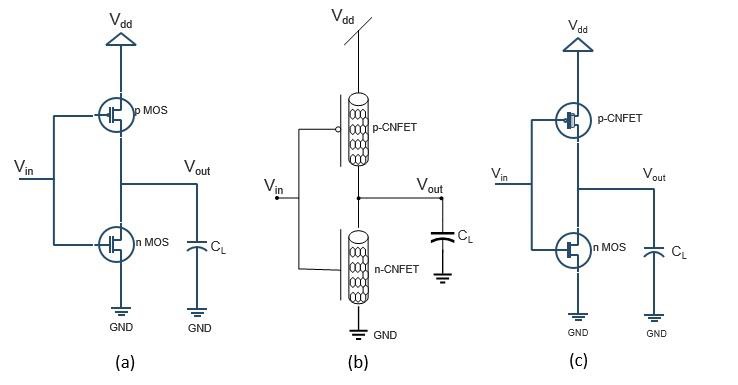}
\caption{\textit{(a) CMOS Technology (b) CNFET Technology (c) Hybrid CMOS-CNFET Technology.}}
\end{center}
\end{figure}

\section{Results and Discussion}
The observed CLA simulation results using hybrid CMOS-CNT are compared with MOS technology and CNT technology results. The comparisons are assessed in terms [10-12] of total voltage source power dissipation, average power consumption, delay, and power-delay product.

\subsection{NP dynamic logic adder}
The parameter values of 8-bit and 16-bit CLA implemented using NP dynamic logic is listed in the table[1]. In terms of total voltage source power dissipation and average power consumption CNFET based CLA gives better performance characteristics than MOSFET. Propagation delay and Power delay product calculations also signifies the CLA performance in CNFET technology.
\begin{table}[h]
\begin{center}
\begin{tabular}{|p{4cm}|p{2.70cm}|p{2.70cm}|p{2.70cm}|p{2.70cm}|}
 \hline
\multirow{2}{*}{Parameter} & \multicolumn{2}{|c|}{8-bit} & \multicolumn{2}{|c|}{16-bit} \\
\cline{2-5}
  &Silicon    &Carbon&  Silicon  &Carbon \\
 \hline
 Total power dissipation (watts)&  85.1609$\mu$  & 33.728n & 132.9407$\mu$ & 61.7694n \\
 \hline
 Average power consumption (watts) &14.42E-05& 2.615E-05 &28.147E-05& 5.665E-05 \\
 \hline
  Propagation Delay (s)& 155.61p& 132.74f &195.02p &385.05f\\
 \hline
 Power delay product(J) &22.438E-15& 3.471E-18 &54.89E-15& 21.81E-18 \\
 \hline
 \end{tabular}
 \caption{\textit{Comparison between CN-MOSFET and SI-MOSFET CLA based on  total power dissipation, average power consumption, propagation delay, power-delay product for 8-bit and 16-bit inputs.}}
\end{center}
\end{table}

The parameter values of 32-bit and 64-bit CLA implemented using NP dynamic logic is listed in the table[2]. In terms of total voltage source power dissipation and average power consumption CNFET based CLA gives better performance characteristics than MOSFET based CLA. Propagation delay, and power delay product calculations also signify the CLA performance in CNFET technology.

\begin{table}[h!]
\begin{center}
\begin{tabular}{|p{4cm}|p{2.70cm}|p{2.70cm}|p{2.70cm}|p{2.70cm}|}
 \hline
\multirow{2}{*}{Parameter} & \multicolumn{2}{|c|}{32-bit} & \multicolumn{2}{|c|}{64-bit} \\
\cline{2-5}
  &Silicon    &Carbon&  Silicon  &Carbon \\
 \hline
 Total power dissipation (watts)&  323.358$\mu$&
76.8077n&
957.513$\mu$&
164.1022n \\
 \hline
 Average power consumption (watts) &43.834E-05&
11.923E-05&
122.50E-05&
15.573E-05 \\
 \hline
  Propagation Delay (s)& 262.36p &
680.36f &
280.43p &
852.07f\\
 \hline
 Power delay product(J) &115E-15&
81.1E-15 &
343.52E-15 &
132.69E-15 \\
 \hline
\end{tabular}
\vspace{0.3cm}
\caption{\textit{Comparison between CN-MOSFET and SI-MOSFET CLA based on  total power dissipation, average power consumption, propagation delay, power-delay product for 32-bit and 64-bit inputs.}}
\end{center}
\end{table}


\subsection{Domino logic adder}
The parameter values of 8-bit and 16-bit CLA implemented using Domino logic is listed in the table [3]. In terms of total voltage source power dissipation and average power consumption CNFET based CLA gives better performance characteristics than MOSFET based CLA. Propagation delay and Power delay product calculations also signifies the CLA performance in CNFET technology.

\begin{table}[h!]
\begin{center}
\begin{tabular}{|p{4cm}|p{2.70cm}|p{2.70cm}|p{2.70cm}|p{2.70cm}|}
 \hline
\multirow{2}{*}{Parameter} & \multicolumn{2}{|c|}{8-bit} & \multicolumn{2}{|c|}{16-bit} \\
\cline{2-5}
  &Silicon    &Carbon&  Silicon  &Carbon \\
 \hline
 Total power dissipation (watts)&  85.3174$\mu$&
35.1642$\mu$&
212.7797$\mu$&
63.5934$\mu$ \\
 \hline
 Average power consumption (watts) &3.0448E-04&
4.6374 E-05 &
3.6008E-04&
5.8339E-05 \\
 \hline
  Propagation Delay (s)& 96.388p7&
201.578f&
102.64p&
339.27f\\
 \hline
 Power delay product(J) &29.348E-15&
9.347E-18&
36.958E-15&
19.792E-18 \\
 \hline
\end{tabular}
\vspace{0.3cm}
\caption{\textit{Comparison between CN-MOSFET and SI-MOSFET CLA based on  total power dissipation, average power consumption, propagation delay, power-delay product for 8-bit and 16-bit inputs.}}
\end{center}
\end{table}

The parameter values of 32-bit and 64-bit CLA implemented using NP dynamic logic is listed in the table [4]. In terms of total voltage source power dissipation and average power consumption CNFET based CLA gives better performance characteristics than MOSFET based CLA. Propagation delay and Power delay product calculations also signifies the CLA performance in CNFET technology.

\begin{table}[h!]
\begin{center}
\begin{tabular}{|p{4cm}|p{2.70cm}|p{2.70cm}|p{2.70cm}|p{2.70cm}|}
 \hline
\multirow{2}{*}{Parameter} & \multicolumn{2}{|c|}{32- bit} & \multicolumn{2}{|c|}{64-bit} \\
\cline{2-5}
  &Silicon    &Carbon&  Silicon  &Carbon \\
 \hline
 Total power dissipation (watts)&  234.8754$\mu$&
79.3373n&
301.5764$\mu$&
153.8624n \\
 \hline
 Average power consumption (watts) &4.4421E-04&
1.2223E-04&
5.8691E-04&
1.9588E-04\\
 \hline
  Propagation Delay (s)& 174.63p&
658.85f&
223.85p&
708.64f\\
 \hline
 Power delay product(J) &77.572E-15&
80.53E-18&
131.37E-15&
138.80E-18 \\
 \hline
\end{tabular}
\vspace{0.3cm}
\caption{\textit{Comparison between CN-MOSFET and SI-MOSFET CLA based on  total power dissipation, average power consumption, propagation delay, power-delay product for 32-bit and 64-bit inputs.}}
\end{center}
\end{table}

\pagebreak

\subsection{NP dynamic logic adder: Revisited}
The comparison of NP dynamic logic CLA and Domino logic CLA in terms of total power dissipation, average power, delay and power-delay product is given in above tables 1,2,3, and 4, respectively. These results observed in stages of 8-bit, 16-bit, 32-bit and 64-bit signifies the performance improvement in NP dynamic CLA than Domino CLA, when CNFET technology is used. Therefore further work is done by considering the NP dynamic logic CLA.

\begin{table}[h!]
\begin{center}
\begin{tabular}{ |p{2.5cm}|p{1.8cm}|p{1.8cm}|p{1.8cm}|p{2cm}|p{1.8cm}|p{1.75cm}| }
 \hline
\multirow{2}{*}{Parameter} & \multicolumn{3}{|c|}{8- bit} & \multicolumn{3}{|c|}{16-bit} \\
\cline{2-7}
  &Silicon    &Carbon& Hybrid& Silicon  &Carbon&Hybrid \\
 \hline
 Total power dissipation (watts)&  85.1609 $\mu$& 33.728n & 31.162n& 132.9407$\mu$& 61.7694n& 59.986n\\
 \hline
 Average power consumption (watts) &14.42E-05&
2.615E-05&
2.147E-05&
28.147E-05&
5.665E-05&
4.263E-05 \\
 \hline
  Propagation Delay (s)& 155.61p&
132.74f&
4.062p&
195.02p&
385.05f&
4.944p\\
 \hline
 Power delay product(J) &22.438E-15 &
3.471E-18&
87.21E-18&
54.89E-15&
21.81E-18&
210.8E-18
 \\
 \hline
\end{tabular}
\vspace{0.3cm}
\caption{\textit{Comparison among Si-, Carbon-, and Hybrid-MOSFET based on  total power dissipation, average power consumption, propagation delay, power-delay product for 8-bit and 16-bit inputs.}}
\end{center}
\end{table}

The parameter values of 8-bit,16-bit, 32-bit and 64-bit CLA implemented using NP dynamic logic is listed in the table[5,6]. In terms of total voltage source power dissipation and average power consumption CNFET based CLA and Hybrid CMOS-CNT based CLA gives equally compatible performance characteristics. But in case of propagation delay CNFET based CLA offers less delay, which gains a low power delay product value than Hybrid CMOS-CNT. In comparison with MOSFET technology, implementation of CLA in CNT and Hybrid CMOS-CNT offer better performance.

\begin{table}[h!]
\begin{center}
\begin{tabular}{ |p{3cm}|p{1.8cm}|p{1.8cm}|p{1.8cm}|p{2cm}|p{1.8cm}|p{1.75cm}| }
 \hline
\multirow{2}{*}{Parameter} & \multicolumn{3}{|c|}{32- bit} & \multicolumn{3}{|c|}{64-bit} \\
\cline{2-7}
  &Silicon    &Carbon& Hybrid& Silicon  &Carbon&Hybrid \\
 \hline
 Total power dissipation (watts)&  323.3528$\mu$&
76.8077n & 77.843n & 957.513$\mu$& 164.1022n& 163.504n\\
 \hline
 Average power consumption (watts) &43.834E-05&
11.923E-05&
10.459E-05&
122.50E-05&
15.573E-05&
14.855E-05 \\
 \hline
  Propagation Delay (s)& 262.36p&
680.36f&
14.337p&
280.43p&
852.07f&
20.137p\\
 \hline
 Power delay product(J) &115E-15&
81.1E-18&
1.49E-15&
343.52E-15&
132.69E-18&
2.99E-15 \\
 \hline
\end{tabular}
\vspace{0.3cm}
\caption{\textit{Comparison among Si-, Carbon-, and Hybrid-MOSFET based on  total power dissipation, average power consumption, propagation delay, power-delay product for 32-bit and 64-bit inputs.}}
\end{center}
\end{table}

\vspace{-1cm}

\section{Summary and Conclusion}
The performance optimization of 64-bit CLA based on CNFET technology signifies the advantage of NP dynamic logic technique in terms of power and delay. Despite the promising progress of CNFETs, the high fabrication cost of CNFETs and fabrication of p-type CNT and n-type CNT on a single chip might cause issues regarding imperfection and variability [8][9]. The challenge facing the Stanford team was that CNTs are predominately p-type semiconductors and there was no easy way to dope these carbon filaments to add n-type characteristics[2].\\

For cost-effective and reliable utilization of CNFETs, and the time-gap reduction in migrating from silicon MOSFET to CNFET technology, the CNFET technology has been required to be combined with low-cost and reliable CMOS technology[13-16]. The CMOS technology is better in switching speed, especially for n-MOS. In this work, the high mobility transport in p-type CNFETs is exploited and combined with high-performance conventional n-type MOSFETs, thereby achieving the best overall performance in a hybrid configuration. Therefore, implementing a hybrid device using CMOS and CNT technology is the best approach to the advancement of nanotechnology field.

\section{References}
\begin{enumerate}
\item \textit{Shimaa I. Sayed, M.M.Abutaleb, Zaki B. Nossair, ``Performance Optimization of Logic Circuits based on Hybrid CMOS and CNFET Design," International Journal of Recent Technology and Engineering (IJRTE) Volume-1.}
 \item \textit{Online Available: https://nano.stanford.edu/stanford-cnfet-model-hspice}.
\item \textit{Yanan Sun and VolkanKursun, ``Carbon Nanotubes Blowing New Life Into NP Dynamic CMOS Circuits" IEEE transactions on circuits and systems—I: regular papers, vol. 61, no. 2, February 2014.}
\item  \textit{J. Deng and H.-S. P.Wong, ``A compact SPICE model for carbon-nanotube field-effect transistors including nonidealities and its application— Part I: Model of the intrinsic channel region," IEEE Trans. Electron Devices, vol. 54, no. 12, pp. 3186–3194, Dec. 2007.}
\item \textit{J. Deng and H.-S. P.Wong, ``A compact SPICE model for carbon-nanotube field-effect transistors including nonidealities and its application— Part II: Full device model and circuit performance benchmarking," IEEE Trans. Electron Devices, vol. 54, no. 12, pp. 3195–3205, Dec. 2007.}
\item \textit{Y.Maheswar, Dr.B.L.Raju, Dr.K.Soundara Rajan ``Modelling and Characterization of CNTFET using Hspice," International Journal of Scientific  Engineering Research, Volume 4, Issue 7, July-2013.
\item A.Naga Lakshmi, Ch.Sirisha ``Performance Optimization of Dynamic and Domino logic Carry Look Ahead Adder using CNTFET in 32nm technology”, IOSR Journal of VLSI and Signal Processing (IOSR-JVSP) Volume 5, Issue 5, pp.30-35, Ver. I (Sep - Oct. 2015).}
\item \textit{Kyung Ki Kim, Yong-Bin Kim and Ken Choi, ``Hybrid CMOS and CNFET Power Gating in Ultralow Voltage Design," IEEE transactions on nanotechnology, vol. 10, no. 6, November 2011.}
\item \textit{S. Lin, Y. Kim, and F. Lombardi, ``CNTFET-based design of ternary logic gates and arithmetic circuits,"
IEEE Trans. Nanotechnol., vol. 10, no. 2, pp. 217–225, Mar. 2011.}
\item \textit{``Online Available: http://ptm.asu.edu/}
\item \textit{``Online Available: http://www.mosis.com/files/scmos/scmos.pdf}
\item \textit{A. P. Chandrakasan, S. Sheng, and R. W. Brodersen, ``Low-power CMOS digital design," IEEE J. Solid-State Circuits, vol. 27, no. 4, pp. 473–484, Apr. 1992.}
\item \textit{T.S. Cho, K.-J. Lee, T. Pan, J. Kong, A.P. Chandrakasan ``Design and Characterization of CNT-CMOS Hybrid System," MTL Annual research report 2007.}
\item \textit{Usmani, Hasan ``Novel hybrid CMOS and CNFET inverting amplifier design for area, power and performance optimization," IEEE Electron Devices and Semiconductor Technology, June-2009.}
\item \textit{Weisheng Zhao, Guillaume Agnus, Vincent Derycke, Ariana Filoramo, Christian Gamra, Jean-
Philippe Bourgoin ``Functional Model of Carbon Nanotube Programmable Resistors for Hybrid Nano/CMOS Circuit Design," Nano-Net, Volume 20 of the series Lecture Notes of the Institute for Computer Sciences, Social Informatics and Telecommunications Engineering.}
\item \textit{Akinwande, Yasuda, Paul, Fujita, Shinobu ``Monolithic integration of CMOS VLSI and CNT for hybrid nanotechnology applications," IEEE Solid-State Device Research Conference, Sept. 2008.}
\end{enumerate}

\end{document}